\documentclass[11pt,twoside]{article}
\usepackage{asp2004}
\usepackage{psfig}
\usepackage{epsf}
\usepackage{graphics}
\usepackage{lscape}
\usepackage{amsmath}
\begin{document}
\title{Progenitors of Type Ia Supernovae: Circumstellar Interaction,
Rotation, and Steady Hydrogen Burning}
\author{Ken'ichi Nomoto, Tomoharu Suzuki, Jinsong Deng, Tatsuhiro
Uenishi and Izumi Hachisu$^*$}
\affil{Department of Astronomy, School of Science, University of Tokyo,
\\7-6-1 Hongo, Bunkyo-ku, Tokyo 113-0033, Japan}
\affil{Department of Earth Science and Astronomy, College of Arts
and Sciences, University of Tokyo, Komaba, Meguro-ku, Tokyo 153-8902,
Japan$^*$}

\begin{abstract}

     Among the important issues in identifying the progenitor system
of Type Ia supernovae (SNe Ia), we focus mostly on circumstellar
interaction in SN 2002ic, and give brief discussion on the
controversial issues of the effects of rotation in merging double
degenerates and steady hydrogen shell burning in accreting white
dwarfs.

     SN 2002ic is a unique supernova which shows the typical spectral
features of SNe Ia near maximum light, but also apparent hydrogen
features that have been absent in SNe Ia.  Based on the hydrodynamical
models of circumstellar interaction in SN Ia, we suggest that
circumstellar medium is aspherical (or highly clumpy) and contains
$\sim 1.3 M_\odot$.  Possible progenitor systems of SN 2002ic are
discussed.

\end{abstract}
\thispagestyle{plain}

\section{Introduction}

     Type Ia supernovae (SNe Ia) are characterized by the lack of
hydrogen and the prominent Si line in their spectra near maximum light
and widely believed to be thermonuclear explosions of mass-accreting
white dwarfs in binary systems.  SNe Ia have been used as a ``standard
candle'' to determine cosmological parameters thanks to their
relatively uniform light curves and spectral evolution.  SNe Ia are
also the major sources of Fe in the galactic and cosmic chemical
evolution.  Despite such importance, the immediate progenitor binary
systems have not been clearly identified yet (e.g., Nomoto et
al. 2000; Livio 2000).

     Recent progress in identifying the progenitor systems includes
the study of circumstellar interaction suggested from high velocity
materials (e.g., Thomas et al. 2004; Gerardy et al. 2004; Mazzali et
al. 2004), especially for SN 2002ic, which has been changed from Type
Ia to Type IIn supernova (Hamuy et al. 2003).  We describe
hydrodynamical models to infer the nature of circumstellar matter and
thus the progenitor (\S2-5).

     The issue of ``{\sl single degenerate vs. double degenerates}''
has long been controversial.  For this we discuss the effects of
rotation in the double degenerate scenario (\S6), and steady hydrogen
shell burning in the single degenerate scenario (\S7).

\section{Circumstellar Medium of Type Ia Supernovae}

     For a model of SN Ia progenitors, Hachisu et al. (1999ab)
proposed a single degenerate model in which the white dwarf blows a
massive and fast wind (up to $10^{-4} M_\odot$~yr$^{-1}$ and
2000~km~s$^{-1}$) and avoids a formation of common envelope when the
mass transfer rate from the normal companion exceeds a critical rate
of $\sim 1 \times 10^{-6}~M_\odot$~yr$^{-1}$ (Nomoto 1982).  Such an
evolutionary phase is dubbed ``accretion wind evolution'' instead of
``common envelope evolution.''  Such a binary can keep its separation
almost unchanged.  The white dwarf can steadily accrete a part of the
transferred matter and eventually reach the Chandrasekhar mass.

     In the strong wind model, the WD winds form a circumstellar
envelope around the binary systems prior to the explosion.  When the
ejecta collide with the circumstellar envelope, X-rays, radio, and
H$\alpha$ lines are expected to be emitted by shock heating.  Attempts
have been made to detect such emissions, but so far no signature of
circumstellar matter has been detected.

     The upper limit set by X-ray observations of SN 1992A is $\dot M
/ v_{10} = (2-3) \times$ $10^{-6} M_\odot$ yr$^{-1}$ (Schlegel,
Finkbeiner, \& Davis 1998).  Radio observations of SN 1986G have
provided the most stringent upper limit to the circumstellar density
as $\dot M / v_{10} = 1 \times 10^{-7} M_\odot$ yr$^{-1}$, where
$v_{10}$ means $v_{10}= v/10$ km s$^{-1}$ (Eck et al. 1995).  This is
still $10-100$ times higher than the density predicted for the white
dwarf winds, because the WD wind velocity is as fast as $\sim 1000$ km
s$^{-1}$.

     For H$\alpha$ emissions, the upper limit of $\dot M / v_{10} =$ 9
$\times$ $10^{-6} M_\odot$ yr$^{-1}$~ has recently been obtained for
SN 2000cx using the ESO/VLT (Lundqvist et al. 2003).

\section {SN 2002ic}

     SN~2002ic was discovered on 2002 November 13~UT at magnitude 18.5
by the Nearby Supernova Factory search (Wood-Vasey et al. 2002).
Hamuy et al. (2003) reported strong Fe III features and a Si II
$\lambda$6355 line in the early-time spectra of SN~2002ic and
classified it as a SN Ia.

    However, strong H$\alpha$ emission was also observed.  The
emission was broad (FWHM $> 1000$ km~s$^{-1}$) suggesting that it was
intrinsic not to an H II region of the host galaxy but to the
supernova (SN). The detection of H$\alpha$ is unprecedented in a SN Ia
(e.g., Branch et al. 1995; Livio 2000).

    Hamuy et al. (2003) suggested that it arose from the
interaction between the SN ejecta and a dense, H-rich circumstellar
medium (CSM), as in Type IIn SNe (SNe IIn).  If this interpretation is
correct, SN~2002ic may be the first SN Ia to show direct evidence of
the circumstellar (CS) gas ejected by the progenitor system,
presenting us with a unique opportunity to explore the CSM around a SN
Ia and the nature of the progenitor system.

\begin{figure}[t]
  \begin{center}
\begin{minipage}[t]{0.9\textwidth}
	\plotone{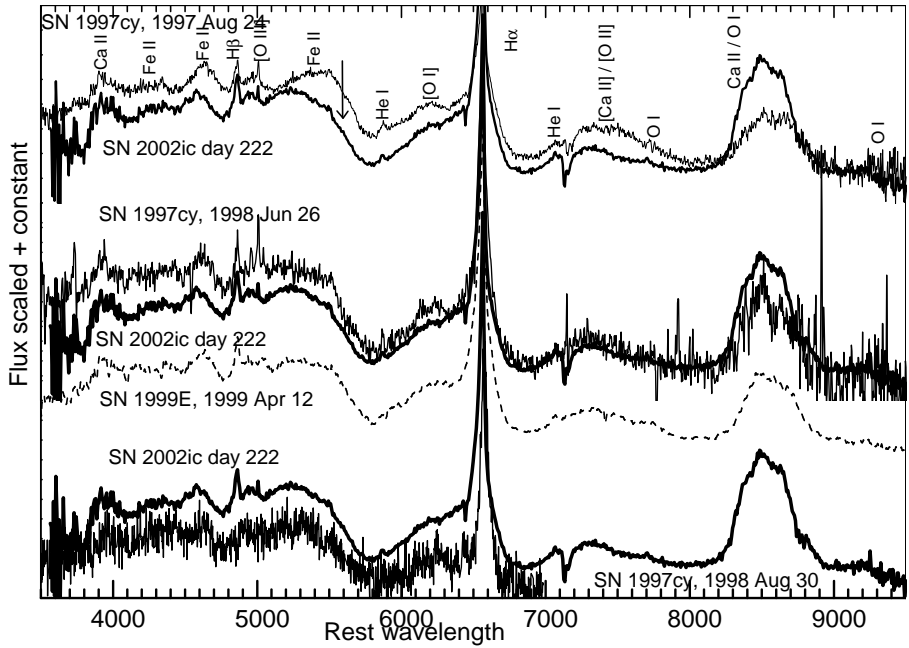}
\end{minipage}
 \end{center}
\caption{Spectral comparison between SNe~2002ic
 (thick lines; Subaru, $\sim 222$ d), 1997cy (thin lines)
 and 1999E (dashed line).
\label{fig1}}
\end{figure}

\subsection {Spectroscopic Features of SN 2002ic}

     The spectrum of SN~2002ic is strikingly similar to those of Type
IIn SNe~1997cy (Turatto et al. 1998) and 1999E (Rigon et al. 2003).
as shown in Figure\ref{fig1}.  
In particular complex line profiles evolve with time,
easily detectable in the prominent H$\alpha$ line.  H$\alpha$
shows at least three components: an unresolved emission on the top of
broader components, which become narrower with time.  At the epoch of
the first observation the broadest component has FWHM $= 12800$ km
s$^{-1}$, and its flux dominates over the intermediate (FWHM $= 4300$
km s$^{-1}$) one. One year after the explosion, the broadest component
has almost disappeared and the intermediate component (which now has
FWHM $= 2000$ km s$^{-1}$) is the most evident spectral feature in the
spectrum.

     SNe~1997cy and 1999E were initially classified as Type IIn
because they showed H$\alpha$ emission. SN~2002ic would also have been
so classified, had it not been discovered at an early epoch.
SN~1997cy ($z=0.063$) is among the most luminous SNe discovered so far
($M_{V}<-20.1$ about maximum light), and SN~1999E is also bright
($M_{V}<-19.5$).  Both SNe~1997cy and 1999E have been suspected to be
spatially and temporally related to a GRB (Germany et al. 2000; Rigon
et al. 2003).  However, both the classification and the associations
with a GRB must now be seen as highly questionable in view of the fact
that their replica, SN~2002ic, appears to have been a genuine SN Ia at
an earlier phase.

\begin{figure}[ht]
\begin{center}
    \begin{minipage}[t]{0.9\textwidth}
      \plotone{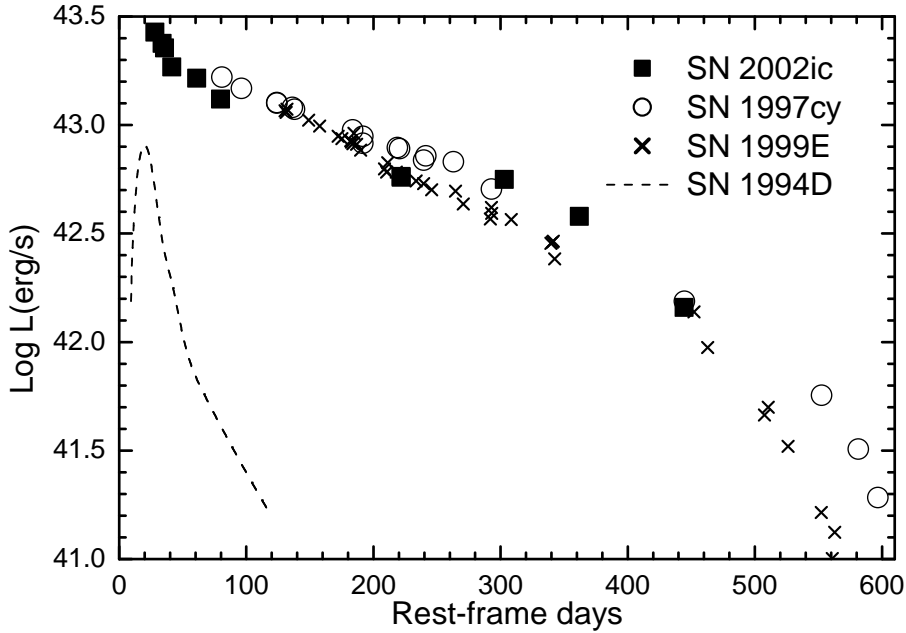}	
    \end{minipage}
\end{center}
\caption{Comparison of the $UBVRI$ bolometric light curves of
SN~2002ic (solid squares) with those of SNe~1997cy (open circles) and
1999E (crosses), and the normal SN Ia 1994D (Contardo et al. 2000;
dashed line).
\label{fig2}}
\end{figure}

\subsection{Observed Light Curve of SN 2002ic}

     The {\sl UVOIR} bolometric light curve of SN 2002ic has been
constructed by Deng et al. (2004) from the available BVRI
photometry and the spectrophotometry (Hamuy et al. 2003; Wang et al. 2004)
as shown in Figure \ref{fig2}.  
To construct the light curve of SN~2002ic, the Subaru spectrum was
integrated. This yielded $L = (5.9\pm 0.6)\times 10^{42}$
ergs~s$^{-1}$, corresponding to $M_{bol}\sim -18.2$, assuming a
distance of 307 Mpc ($H_0$ = 65 km~s$^{-1}$~Mpc$^{-1}$).

     The light curve of SN 2002ic is brighter at maximum and declines
much more slowly than typical SNe Ia (Hamuy et al. 2003).  The late time
light curve of most SNe is powered by the radioactive decay of
$^{56}$Co to $^{56}$Fe.  The decline of SN 2002ic is even slower than the
Co decay rate, which indicates the presence of another source of
energy.

     In fact the overall light curve of SN 2002ic resembles Type IIn
SN 1997cy (Turatto et al. 1998), suggesting circumstellar interaction
for the energy source.  Assuming $A_V=0.00$ for the galactic
extinction (NED) SN 2002ic is a factor of 1.3 dimmer than SN 1997cy.
We use $UBVRI$ bolometric light curves of SNe~1997cy and 1999E for
comparison (Turatto et al. 1998; Rigon et al. 2003) (with a phase
computed from their assumed GRB counterparts, which cannot be greatly
in error even if, as seems likely, the GRB associations are
incorrect).  The $UBVRI$ bolometric light curves of the three SNe are
also very similar (see Figure\ref{fig2}).

\begin{figure}
\begin{center}
	\begin{minipage}[t]{0.9\textwidth}% !Here input the size you like!
%		\plotone{lumi.eps}
		\plotone{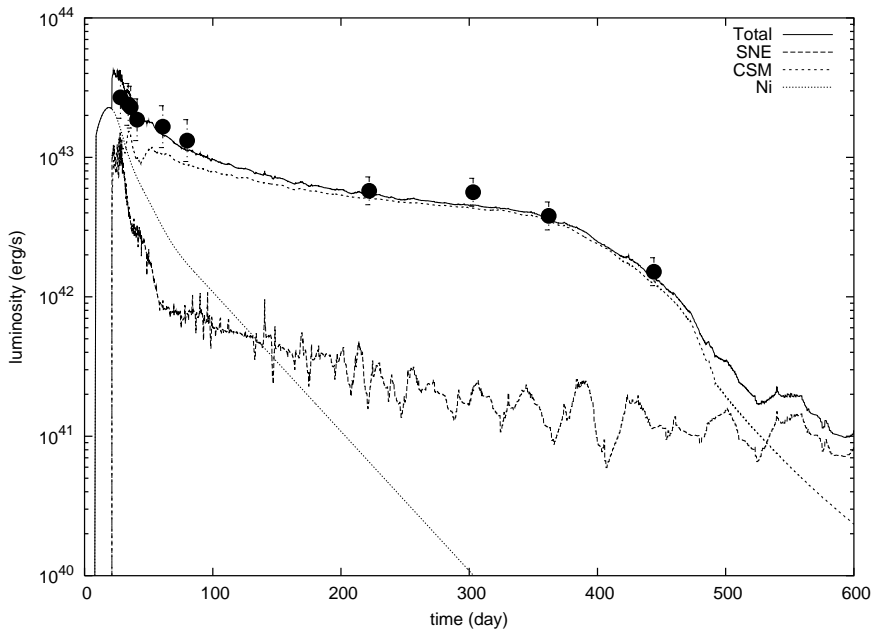}
	\end{minipage}
\caption{Calculated light curve with observed data points
(filled circles: Deng et al. 2004).
\label{lumi}}
\end{center}
\end{figure}

\section{Circumstellar Interaction Models}

     We calculated the interaction between the expanding ejecta and
CSM.  For the supernova ejecta, we used the the carbon deflagration
model W7 (Nomoto, Thielemann, \& Yokoi 1984); its kinetic energy is $E
= 1.3 \times$ $10^{51}$ erg.  For CSM we assumed the power-law density
distribution:

\begin{equation}
\rho = \rho_0(r/R_0)^{-n}  \mathrm{g \ cm^{-3}}
\end{equation}
where the parameters are the radius ($R_0$) and density ($\rho_0$) of
the point where the ejecta and CSM start interacting, and the index
($n$) of the density distribution of CSM.  These parameters are
constrained from comparison with the observed light curve.  

     When the expanding ejecta interacts with CSM, the interaction
creates the forward shock which is propagating through the CSM and the
reverse shock which is propagating through the ejecta (propagating
backwards in Lagrangian scheme).

     Shocked matter is heated to $T \sim 10^{7}$ K for the reverse shock
and $T \sim 10^{9}$ K for the forward shock. Both shocked regions emit
thermal X-rays. For the reverse shock, because of relatively high
densities in the ejecta, cooling time scale is shorter than shock
propagation so that the shocked ejecta soon forms a dense cool shell
(Suzuki \& Nomoto 1995). This dense cool shell absorbs the X-ray and
re-emits in UV-optical. This re-emitted photons are observed.  We
assume that a half of the X-rays emitted in the reverse-shocked ejecta
is lost into the supernova center, and that the other half is
transfered outwardly through the cooling shell.  We also assume that a
half of the X-rays emitted in the CSM is transfered inwardly to be
absorbed by the cooling shell.  We take into account the change in
time of the column density of the cooling shell to evaluate the X-rays
absorbed by the shell and the optical luminosity.

\begin{figure}
\begin{center}
	\begin{minipage}[t]{0.9\textwidth}%!Here input the size you like!
%		\plotone{velo.eps}
		\plotone{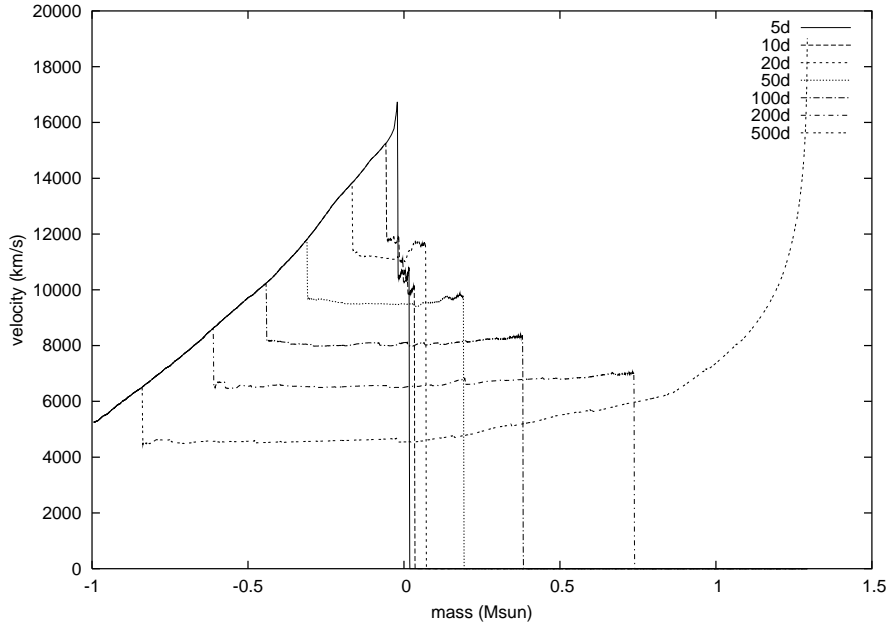}
	\end{minipage}
\caption{Velocity profile in the interacting ejecta and CSM.
\label{velo}}
\end{center}
\end{figure}

     Figure \ref{lumi} shows the successful model for the light curve
of SN 2002ic.  The model light curve includes the radioactive decay
using the bolometric light curve of SN 1991T.  Recently, Wood-Vasey et
al. (2004) reported the presence of a $\sim 1.7 \times$ $10^{15}$ cm
gap from their analysis of the light curve.  Therefore, we set $R_0 =
1.7 \times$ $10^{15}$ cm.  For inner CSM of 1.1 $M_\odot$, $\rho_0 =
1.6 \times$ $10^{-15}$ g cm$^{-3}$ and $n = 1.8$.  For the wind
velocity of $v_w =$ 10 km s$^{-1}$, these parameters correspond to the
mass-loss rate of $\dot{M} \sim 1 \times$ $10^{-3}$ $M_\odot$
yr$^{-1}$.  In our initial model, the CSM is extend to $3.3\times$
$10^{16}$ cm, and this means that the super wind lasted for $\sim
10^3$ yr.  In the early phase, the model with $n = 2.0$ (steady mass
loss) declines too fast to be compatible with the observation.  This
implies that CSM around the SN was created by unsteady mass loss of
the progenitor system.

     After day $\sim$ 350, the light curve starts declining.  To
reproduce the declining part of the light curve, we add the outer CSM
of 0.2 $M_\odot$~ where the density declines sharply as $n = 6$.  This
implies that the total mass of CSM is $\sim 1.3 M_\odot$.

     The mass of CSM is much smaller than $\sim 4.7 M_\odot$ estimated
in Nomoto et al. (2004), where we adopted solar chemical composition
for the ejecta.  As also reported in Chugai et al. (2004), a realistic
chemical composition of the SN Ia ejecta results in more efficient
cooling in the reverse shock and more efficient absorption of the
X-rays in the supernova ejecta.  These effects lead to the lower
density and smaller total mass of CSM.

     We note in Figure \ref{velo} that the velocity of the ejecta is
$\la$ 8,000 km s$^{-1}$ and too low for the observed broad component
($\ge$ 10,000 km s$^{-1}$).  In fact, in order to produce high
enough luminosity to explain the light curve, such a strong
interaction between the ejecta and CSM should occur.

     To reproduce both the light curve and the observed velocity of SN
2002ic, CSM needs to be aspherical.  Suppose the CSM is aspherical
consisting of a dense region and a thin region.  The expanding ejecta
interacting strongly with the dense region can produce high enough
luminosity to explain the light curve.  On the other hand, the ejecta
interacting with the thin region can expand still fast enough to be
consistent with the observed velocities.

\section{Possible Progenitors of SN 2002ic}

There are two possible progenitor scenarios for SN 2002ic.  One is the
explosion of the C+O core of the massive AGB star (SN I+1/2), where
the wind from the AGB star formed the CSM.  The other is the explosion
of the white dwarf in a close binary blowing wind to create the dense
CSM (e.g., Livio \& Riess 2003; Chugai \& Yungelson 2004).

\subsection {Type I+1/2 Supernovae in AGB Stars}

Single star scenario is the explosion of the massive AGB star whose
C+O core becomes very close to the Chandrasekhar mass.  Before
explosion, mass loss (super-wind) from the star creates a dense CSM.
The C+O core explodes, which is called Type I+1/2 supernova, and
interacts with CSM.

To make this scenario possible, the metallicity of the system should
be low because low mass loss rate is necessary for the C+O to grow to
reach the Chandrasekhar mass before the envelope is completely lost.
Under the solar metallicity, SN I+1/2 have never been observed.
Therefore, we can explain the rarity of SN 2002ic-like event assuming
that only narrow mass range of AGB stars can explode as SNe in low
metal environment.

Aspherical CSM is not unexpected for stars approaching the end of the
AGB.  A pre-existing clumpy disk was also suggested by Wang et al.
(2004), based on spectropolarimetry.

\subsection {White Dwarf Winds}

Binary star scenario is the explosion of the accreting C+O white dwarf
(same as typical SNe Ia). However, the companion star is massive and
the WD blows large amount of accreting gas as accretion wind to create
the dense CSM.  In this scenario, the rarity is can be attributed to
the fewness of the companion star massive enough to produce the quite
massive CSM.

As a progenitor of SN 2002ic, we need a CSM of $\sim 1.3~M_\odot$.
Such a massive CSM is possible only when the donor is as massive as $4
- 5 M_\odot$.  For the model consisting of a WD and a main-sequence
(MS) companion (Hachisu et al. 1999b), the mass transfer rate from
such a massive MS companion reaches $\sim 1 \times 10^{-4}
M_\odot$~yr$^{-1}$. Then the WD blows a wind of $\sim 1 \times 10^{-4}
M_\odot$~yr$^{-1}$ and the mass stripping rate becomes several times
larger than the WD wind mass loss rate (Hachisu \& Kato 2003).

For the symbiotic model consisting of a white dwarf and a red giant or
AGB star, the wind mass loss rate during the super-wind phase can also
reach $\sim 1 \times 10^{-4} M_\odot$~yr$^{-1}$. In symbiotic stars,
the mass capture efficiency by the WD is observationally estimated to
be as small as one or a few percent. Therefore, only when a large part
of the red giant wind or AGB super-wind is captured by the white dwarf,
the white dwarf can blow a very massive wind of $\sim 1 \times 10^{-5}
M_\odot$~yr$^{-1}$ or more.  Then, the mass stripping rate from the red
giant or AGB star also reaches several times $10^{-4}
M_\odot$~yr$^{-1}$ or more.

Examples of the accretion wind evolution are identified as transient
supersoft X-ray sources, i.e., the LMC supersoft X-ray source
RX~J0513.9$-$6951 and its Galactic counterpart V~Sge (Hachisu \& Kato
2003).  Especially in V~Sge, a very massive wind of $\sim 1 \times
10^{-5} M_\odot$~yr$^{-1}$ has been observationally suggested by the
detection of radio.  Furthermore, the white dwarf wind collides with
the companion and strips heavily off its surface matter.  This
stripping rate reaches a few or several times the wind mass loss rate
of the white dwarf, i.e., $\sim 1 \times 10^{-4} M_\odot$~yr$^{-1}$ or
more (Hachisu \& Kato 2003). The matter stripped off has a much lower
velocity than the white dwarf wind itself and forms an excretion disk
around the binary.  Thus the model predict the coexistence of a fast
white dwarf wind blowing mainly in the pole direction and a massive
disk or a torus around the binary.  Deng et al. (2004) propose a new
classification, Type IIa SNe, for these events.

\section{Rotation and Merging White Dwarfs}

The issue of ``single degenerate vs. double degenerates'' has long
been controversial.  Nomoto \& Iben (1985) and Saio \& Nomoto (1998)
have simulated the merging of double WDs in 1D and shown that the
rapidly accreting WDs undergo off-center carbon ignition if $\dot M$
$>$ 2 $\times$ 10$^{-6}$ $M_\odot$ yr$^{-1}$ because of compressional
heating.  Afterwards carbon flame propagates inward through the center
and converts C+O into O+Ne+Mg.  Then the final outcome is most likely
accretion-induced collapse rather than SNe Ia.

Recently Piersanti et al. (2003a,b) calculated the evolution of WDs
with rotation in 1D approximation.  They assumed that all the angular
momentum associated with the accreted matter was brought into the
white dwarf.  They claim that the combined effects of accretion and
rotation induce expansion to make the surface zone gravitationally
unbound and hence suppresses further accretion in the double white
dwarf merger.  They argued that the above effect makes the accretion
rate smaller than the critical value for the occurrence of the
off-center carbon ignition, and hence the white dwarf can grow up to
the Chandrasekhar mass to become a SN~Ia.  However, they did not take
into account the backward transport of angular momentum to the disk.

Saio \& Nomoto (2004) have also calculated the accretion of C+O onto
the C+O WD with rotation for various timescale of angular momentum
transport in 1D approximation.  The outermost layer of the accreting
WD quickly reaches the critical rotation.  Afterwards, the angular
momentum is transported backward to disk and accretion continues
(Paczy\'nski 1991; Popham \& Narayan 1991).  For $\dot M$ $>$ 3
$\times$ 10$^{-6}$ $M_\odot$ yr$^{-1}$, off-center carbon burning is
ignited prior to the central C-ignition.  The difference in the total
mass at the ignition between rotating and non-rotating models can be
as large as $\sim 0.1M_\odot$, depending on the assumed turbulent
viscosity and the accretion rate.  Thus the lifting effect of rotation
increases the critical accretion rate for the occurrence of off-center
C-ignition by a factor of $\sim$ 1.5 compared with the non-rotating
case, but the basic conclusion is the same as non-rotating case, i.e.,
the accretion-induced collapse is the most likely outcome in the
double degenerates scenario (Saio \& Nomoto 1998).

For comparison, we note that Yoon \& Langer (2004) also computed white
dwarf models accreting CO-rich matter, taking into account the effect
of rotation.  The accretion rates they considered are lower than those
adopted in the above calculations, thus leading to the C-ignition.
They assumed that the matter accretes onto the white dwarf without
bringing angular momentum when the rotation velocity at the surface of
the white dwarf exceeds the Keplerian velocity.  Under their
assumption the total angular momentum of the white dwarf never
decreases and is higher than Saio \& Nomoto (2004) models for a given
mass.

\section{Surface Hydrogen Burning Models}

For the single degenerate scenario, steady hydrogen-shell burning
(Sienkiewicz 1975), which consumes hydrogen at the same rate as the
white dwarf accretes, is a major evolutionary route to Type Ia
supernovae (Nomoto 1982).

Recently, Starrfield et al. (2004) presented ''{\sl surface hydrogen
burning models}'' where steady and stable hydrogen burning occurs to
increase the white dwarf mass to an SN Ia for the accretion rate in
the range of $10^{-9} - 10^{-6} M_\odot$ yr$^{-1}$.  This range is
much wider than previous studies have found (Sienkiewics 1980).

Saio et al. (2004) have revisited the properties of white dwarfs
accreting hydrogen-rich matter by constructing steady-state models and
examining thermal stability of those models.  Saio et al. (2004) have
confirmed the results of Sienkiewicz (1975, 1980) and concluded that
the steady ``{\sl surface hydrogen burning}'' of Starrfield et
al. (2004) is an artifact which arose from the lack of resolution for
the envelope structure of their model.

In the surface zone (sz), the pressure at the burning shell is
obtained as $P_{\rm sz} = GM \Delta M/4 \pi R^4$, which is $9 \times
10^{19}$ dyn cm$^{-2}$ for $M = 1.35 M_\odot$, $R=2400$ km, and
$\Delta M = 1 \times 10^{-5} M_\odot$, where $10^{-5} M_\odot$ is the
mesh size of their ``{\sl surface zone}'' (Starrfield et al. 2004).
Because of such a high pressure, the temperature in the surface zone
is as high as $T_{\rm sz} \sim 5 \times 10^8$ K.  When hydrogen-rich
matter is accreted in the surface zone, $T_{\rm sz}$ is high enough to
burn hydrogen and helium immediately, and the nuclear energy
generation rate is determined by the mass accretion rate.  Such a
``steady'' burning is just due to the too coarse mesh size of their
surface zone.

Actual hydrogen-burning occurs in such a superficial layer as $\sim
10^{-7} M_\odot$.  Then ``surface'' hydrogen burning should be
unstable to flash for $\dot M < 10^{-7} M_\odot$ yr$^{-1}$
(Sienkiewicz 1980; Saio et al. 2004). 

\bigskip
\acknowledgments

We would like to thank H. Saio and M. Kato for discussion on rotation
and surface hydrogen burning.  This work was supported in part by the
Grant-in-Aid for Scientific Research (15204010, 16042201, 16540229)
and the 21st Century COE Program (QUEST) from Japan Society for the
Promotion of Science, and the Ministry of Education, Culture, Sports,
Science, and Technology of Japan.

%\begin{thebibliography}{99}

\end{document}